\documentclass[12pt,a4paper]{article}
\begin{document}
\large

\newpage
\begin{center}
{\bf AN AXIAL-VECTOR NATURE OF A NEUTRINO WITH AN ELECTROWEAK MASS}
\end{center}
\vspace{0.5cm}
\begin{center}
{\bf Rasulkhozha S. Sharafiddinov}
\end{center}
\vspace{0.5cm}
\begin{center}
{\bf Institute of Nuclear Physics, Uzbekistan Academy of Sciences,
Tashkent, 100214 Ulugbek, Uzbekistan}
\end{center}
\vspace{0.5cm}

A classification of elementary particles with respect to C-operation
admits the existence of truly neutral types of fermions. Among them one
can find both a Dirac and a Majorana neutrinos of an electroweak nature.
Their mass includes the electric and weak parts, in the presence of which
a neutrino has the anapole charge, charge radius and electric dipole moment.
They constitute the paraneutrino of true neutrality, for example, at the
neutrino interaction with a spinless nucleus of an axial-vector current.
We derive the united equations which relate the structural components of
mass to anapole, charge radius and electric dipole of each truly neutral
neutrino at the level of flavour symmetry. Such a principle can explain the
C-noninvariant nature of neutrinos and fields in the framework of constancy
law of the size implied from the multiplication of a weak mass of the C-odd
neutrino by its electric mass. From this point of view, all neutrinos of
C-antisymmetricality regardless of the difference in masses of an axial-vector
character, possess the same anapole with his radius as well as the identical
electric dipole. Their analysis together with earlier measurements of an
electric dipole moment of neutrino gives the right to accept not only the
existence of truly neutral types of Dirac neutrinos and fields but also the
availability of mass structure in them as the one of experimentally
established facts.

\newpage
\begin{center}
{\bf 1. Introduction}
\end{center}

Our notions about neutrinos of the Dirac \cite{1} and Majorana \cite{2}
nature can be based on the fact that the same particle has no simultaneously
both a C-invariant and a C-noninvariant interaction. This principle, however,
encounters many problems which require the classification of currents with
respect to C-operation.

Such a procedure suggested by the author \cite{3} admits the existence of
truly neutral types of leptons and fields. Their nature has been created so
that to any C-antisymmetrical $(l^{A}=C\overline{l}^{A})$ lepton corresponds
a kind of the Dirac $(\nu_{l}^{A}=C{\bar \nu_{l}^{A}})$ neutrino. They
constitute herewith the united families of fermions of an axial-vector
$(A)$ character:
\begin{equation}
\pmatrix{\nu_{e}^{A}\cr e^{A}}_{L},
(\nu_{e}^{A}, \, \, \, \, e^{A})_{R}, \, \, \, \,
\pmatrix{\nu_{\mu}^{A}\cr \mu^{A}}_{L},
(\nu_{\mu}^{A}, \, \, \, \, \mu^{A})_{R}, \, \, \, \,
\pmatrix{\nu_{\tau}^{A}\cr \tau^{A}}_{L},
(\nu_{\tau}^{A}, \, \, \, \, \tau^{A})_{R}, ...,
\label{1}
\end{equation}
\begin{equation}
\pmatrix{{\bar \nu_{e}^{A}}\cr {\bar e^{A}}}_{R},
({\bar \nu_{e}^{A}}, \, \, \, \, {\bar e^{A}})_{L}, \, \, \, \,
\pmatrix{{\bar \nu_{\mu}^{A}}\cr {\bar \mu^{A}}}_{R},
({\bar \nu_{\mu}^{A}}, \, \, \, \, {\bar \mu^{A}})_{L}, \, \, \, \,
\pmatrix{{\bar \nu_{\tau}^{A}}\cr {\bar \tau^{A}}}_{R},
({\bar \nu_{\tau}^{A}}, \, \, \, \, {\bar \tau^{A}})_{L}, ....
\label{2}
\end{equation}

Unlike the corresponding type of lepton, each $\nu_{l}^{A}({\bar \nu_{l}^{A}})$
neutrino has a self Majorana $\nu_{M}^{A}({\bar \nu_{M}^{A}})$ neutrino. These
pairs are united in the purely neutrino families \cite{4}:
\begin{equation}
\pmatrix{\nu_{e}^{A}\cr \nu_{1}^{A}}_{L},
(\nu_{e}^{A}, \, \, \, \, \nu_{1}^{A})_{R}, \, \, \, \,
\pmatrix{\nu_{\mu}^{A}\cr \nu_{2}^{A}}_{L},
(\nu_{\mu}^{A}, \, \, \, \, \nu_{2}^{A})_{R}, \, \, \, \,
\pmatrix{\nu_{\tau}^{A}\cr \nu_{3}^{A}}_{L},
(\nu_{\tau}^{A}, \, \, \, \, \nu_{3}^{A})_{R}, ...,
\label{3}
\end{equation}
\begin{equation}
\pmatrix{{\bar \nu_{e}^{A}}\cr {\bar \nu_{1}^{A}}}_{R},
({\bar \nu_{e}^{A}}, \, \, \, \, {\bar \nu_{1}^{A}})_{L}, \, \, \, \,
\pmatrix{{\bar \nu_{\mu}^{A}}\cr {\bar \nu_{2}^{A}}}_{R},
({\bar \nu_{\mu}^{A}}, \, \, \, \, {\bar \nu_{2}^{A}})_{L}, \, \, \, \,
\pmatrix{{\bar \nu_{\tau}^{A}}\cr {\bar \nu_{3}^{A}}}_{R},
({\bar \nu_{\tau}^{A}}, \, \, \, \, {\bar \nu_{3}^{A}})_{L}, ....
\label{4}
\end{equation}

Any family distinguishes from others by a true flavour. Each particle can
therefore be characterized by the three $(l^{A}=e^{A},$ $\mu^{A},$ $\tau^{A})$
true flavours \cite{5}:
\begin{equation}
T_{l^{A}}=\left\{
{\begin{array}{l}
{-1\quad \; for\quad \; l^{A}_{L}, \, \, \, \, \, l^{A}_{R}, \, \, \, \,
\nu_{lL}^{A}, \, \, \, \, \nu_{lR}^{A}, \, \, \, \,
\nu_{ML}^{A}, \, \, \, \, \nu_{MR}^{A},}\\
{+1\quad \; for\quad \; \overline{l}^{A}_{R}, \, \, \, \,
\overline{l}^{A}_{L}, \, \, \, \,
{\bar \nu_{lR}^{A}}, \, \, \, \, {\bar \nu_{lL}^{A}}, \, \, \, \,
{\bar \nu_{MR}^{A}}, \, \, \, \, {\bar \nu_{ML}^{A}},}\\
{\, \, \, \, 0\quad \; for\quad \; remaining \, \, \, \, particles.}\\
\end{array}}\right.
\label{5}
\end{equation}

The legality of conservation of all types of true flavours
\begin{equation}
T_{l^{A}}=const
\label{6}
\end{equation}
or full true number
\begin{equation}
T_{e^{A}}+T_{\mu^{A}}+T_{\tau^{A}}=const
\label{7}
\end{equation}
follows from the fact that the unidenticality of an axial-vector
lepton and its antiparticle is the consequence of their CP-symmetrical
$(l^{A}=-CP\overline{l}^{A})$ nature \cite{5}. This picture has important
aspects in the field of emission of the corresponding types of gauge bosons.

One of them states that the interaction of C-noninvariant neutrino with
virtual axial-vector photon \cite{6} can be expressed in the form
of $A_{\nu_{l}}$ current
\begin{equation}
j_{em}^{A_{\nu_{l}}}=
\overline{u}(p',s')\gamma_{5}[\gamma_{\mu}G_{1\nu_{l}^{A}}(q^{2})-
i\sigma_{\mu\lambda}q_{\lambda}G_{2\nu_{l}^{A}}(q^{2})]u(p,s)
\label{8}
\end{equation}
where $\sigma_{\mu\lambda}=[\gamma_{\mu},\gamma_{\lambda}]/2,$ $l=D=e,$
$\mu,$ $\tau$ or $M=1,$ $2,$ $3,$ $p(s)$ and $p'(s')$ imply the four-momentum
(helicities) of neutrino before and after the emission of an axial-vector
nature, $q=p-p',$ $G_{1\nu_{l}^{A}}$ and $G_{2\nu_{l}^{A}}$ characterize,
respectively, the Dirac and Pauli parts of this current \cite{7}.

One can define their structure by the following manner \cite{8}:
\begin{equation}
G_{i\nu_{l}^{A}}(q^{2})=g_{i\nu_{l}^{A}}(0)+R_{i\nu_{l}^{A}}(q^{2})+...,
\label{9}
\end{equation}
in which $g_{i\nu_{l}^{A}}(0)$ give the corresponding sizes of
the anapole \cite{9} and electric dipole \cite{10}. The terms
$R_{i\nu_{l}^{A}}(q^{2})$ there arise in the neutrino axial-vector
radius dependence.

Insofar as $R_{1\nu_{l}^{A}}(q^{2})$ is concerned, it describes a connection
between the anapole $r_{\nu_{l}^{A}}$ radius and field of emission:
$R_{1\nu_{l}^{A}}(q^{2})=(q^{2}/6)<r^{2}_{\nu_{l}^{A}}>.$

According to the earlier presentations about the nature of $A_{\nu_{l}}$
current, $g_{1\nu_{l}^{A}}(0)$ and $g_{2\nu_{l}^{A}}(0)$ constitute,
respectively, its CP-even and CP-odd components of the different T-parity
with the same P-unparity. But we can say that in this situation, there are some
nontrivial contradictions between the anapole and electric dipole. The point is
that to each type of charge corresponds a kind of the dipole \cite{11}. At the
same time, the anapole itself come forwards in a system as an electric charge
of C-odd character \cite{7}. Thereby, it expresses the C-noninvariance of an
electric dipole moment. Therefore, the T-antisymmetricality of the latter
implies the T-unparity of the earlier T-symmetrical anapole even at the absence
of CPT-symmetry of all types of CP-symmetrical $A_{\nu_{l}}$ currents \cite{3}.

On the other hand, as known, the existence of the anapole in neutrino is
incompatible with gauge invariance, from conservation of which would follow
its equality to zero. However, at the availability of mass structure of gauge
symmetry \cite{4}, such a possibility is realized only in the case when a
particle possesses no a self inertial mass.

Another characteristic moment is the mass-charge duality \cite{12}. From its
point of view, any type of charge says in favor of a kind of inertial mass.
An electroweakly charged neutrino can therefore have simultaneously both
a weak \cite{13} and an electric masses.

Furthermore, if a neutrino is of C-invariant fermions, it possesses the
mass of vector character \cite{7}. In contrast to this, the mass of the
C-odd neutrino has an axial-vector nature.

So, it is seen that the mass $m_{\nu_{l}^{A}}$ and charge $e_{\nu_{l}^{A}}$
in all C-noninvariant types of neutrinos with an electroweak behavior are
strictly their electroweakly united $(EW)$ mass and charge
\begin{equation}
m_{\nu_{l}^{A}}=
m_{\nu_{l}^{A}}^{EW}=m_{\nu_{l}^{A}}^{E}+m_{\nu_{l}^{A}}^{W},
\label{10}
\end{equation}
\begin{equation}
e_{\nu_{l}^{A}}=
e_{\nu_{l}^{A}}^{EW}=e_{\nu_{l}^{A}}^{E}+e_{\nu_{l}^{A}}^{W}
\label{11}
\end{equation}
including both an electric $(E)$ and a weak $(W)$ components.

They establish an intraneutrino harmony \cite{14} of the different types
of forces responsible for steadiness of the anapole charge distribution
in neutrino of true neutrality. Such a role of compound structure of the
neutrino mass may serve as an indication in favour of a certain latent
regularity of its axial-vector interaction. This, however, requires the
explanation from the point of view of the discussed types of currents.

Therefore, we investigate here the nature of truly neutral neutrinos of Dirac
$(l=D)$ and Majorana $(l=M)$ types at the elastic scattering in the field of
a spinless nucleus going at the expense of Coulomb $(E)$ and weak $(W)$ masses,
and also of the anapole $g_{1\nu_{l}^{A}}(0)$ charge, charge $r_{\nu_{l}^{A}}$
radius and electric dipole $g_{2\nu_{l}^{A}}(0)$ moment of longitudinal
polarized particles of axial-vector weak neutral $A_{\nu_{l}}$ currents.

For our purposes, it is desirable to choose only the one of spinless
nuclei, in which neutrons and protons are strictly the fermions of true
neutrality \cite{5}. At the same time, all reasoning of a given work refer
to those neutrinos and nuclei, among which no elementary objects
of vector character.

Starting from such a presentation, we establish here an explicit mass
structure dependence of axial-vector currents of neutrinos of a different
nature. Its analysis together with measured value \cite{15,16} of an electric
dipole moment of neutrino shows clearly that the existence of truly neutral
types of Dirac neutrinos and gauge fields as well as the availability of
mass structure in them are practically not excluded.

\begin{center}
{\bf 2. Unity of neutrino axial-vector interaction
structural components}
\end{center}

The above noted electroweak transitions in the limit of one-boson exchange
can be expressed by the two matrix elements \cite{8}:
$$M^{E}_{fi}=\frac{4\pi\alpha}{q_{E}^{2}}\overline{u}(p_{E}',s')\gamma_{5}
\{\gamma_{\mu}[g_{1\nu_{l}^{A}}^{E}(0)+
\frac{1}{6}q_{E}^{2}<r^{2}_{\nu_{l}^{A}}>_{E}]-$$
\begin{equation}
-i\sigma_{\mu\lambda}q_{\lambda E}g_{2\nu_{l}^{A}}^{E}(0)\}
u(p_{E},s)<f|J_{\mu}^{\gamma}(q_{E})|i>,
\label{12}
\end{equation}
\begin{equation}
M^{W}_{fi}=
-\frac{G_{F}}{\sqrt{2}}\overline{u}(p_{W}',s')\gamma_{\mu}\gamma_{5}
g_{A_{\nu_{l}}}^{*}u(p_{W},s)<f|J_{\mu}^{Z}(q_{W})|i>.
\label{13}
\end{equation}

Among them $\nu_{l}^{A}=\nu_{lL,R}^{A}({\bar \nu_{lR,L}}^{A}),$
$q_{E}=p_{E}-p_{E}',$ $q_{W}=p_{W}-p_{W}',$ $p_{E}(p_{W})$ and $p_{E}'(p_{W}')$
characterize in the Coulomb (weak) scattering the four-momentum of incoming and
outgoing neutrinos, $J_{\mu}^{\gamma}$ and $J_{\mu}^{Z}$ correspond to the
target nucleus currents at the interactions with photon and $Z$-boson,
$g_{A_{\nu_{l}}}^{*}$ distinguishes from $g_{A_{\nu_{l}}},$ namely, from
the constant of a neutrino axial-vector weak neutral current by a multiplier
$(1/\sin\theta_{W})$ which describes a situation when
$e_{\nu_{l}^{A}}^{E}=1$ and
\begin{equation}
e_{\nu_{l}^{A}}^{E}=e_{\nu_{l}^{A}}^{W}\sin\theta_{W}.
\label{14}
\end{equation}

The index $E$ in $g_{i\nu_{l}^{A}}^{E}$ and $<r^{2}_{\nu_{l}^{A}}>_{E}$ implies
their neutrino electric mass dependence. It should also be added \cite{17} that
the interaction with field of emission of a neutrino of the Majorana type is
stronger $(g_{A_{\nu_{M}}}=2g_{A_{\nu_{D}}})$ than of a Dirac neutrino
of true neutrality.

In the presence of both the neutrino electric and weak masses, the investigated
processes become possible owing to the mixedly interference $(I)$ interaction
$$ReM^{E}_{fi}M^{*W}_{fi}=
-\frac{4\pi\alpha G_{F}}{\sqrt{2}q_{I}^{2}}
Re\Lambda_{I}\Lambda_{I}'\gamma_{5}\{\gamma_{\mu}
[g_{1\nu_{l}^{A}}^{I}(0)+$$
$$+\frac{1}{6}q_{I}^{2}<r^{2}_{\nu_{l}^{A}}>_{I}]-$$
\begin{equation}
-i\sigma_{\mu\lambda}q_{\lambda I}g_{2\nu_{l}^{A}}^{I}(0)\}
\gamma_{\mu}\gamma_{5}g_{A_{\nu_{l}}}^{*}
J_{\mu}^{\gamma}(q_{I})J_{\mu}^{Z}(q_{I}).
\label{15}
\end{equation}

The availability of $g_{i\nu_{l}^{A}}^{I}$ and $<r^{2}_{\nu_{l}^{A}}>_{I}$
in it indicates to the existence of a connection of $A_{\nu_{l}}$ currents
of a neutrino with the structural parts of its mass and charge.
We see in addition that
$$q_{I}=p_{I}-p_{I}',$$
$$\Lambda_{I}=u(p_{I},s)\overline{u}(p_{I},s),$$
$$\Lambda_{I}'=u(p_{I}',s')\overline{u}(p_{I}',s').$$
Here $p_{I}$ and $p_{I}'$ denote the four-momentum of a fermion before and
after the electroweak emission of an interference character.

They show that between the interference $(I)$ mass of the neutrino and its
all $(EW)$ rest mass takes place the inequality. Such an unidenticality of
both types of masses can explain the difference in the neutrino electric $(E)$
and weak $(W)$ masses. This distinction is, as well as (\ref{14}), responsible
for an electroweak unification at the more fundamental dynamical level.

In the case of spinless nuclei of the C-noninvariant nature and of longitudinal
polarized neutrinos of true neutrality, the studied process cross section with
the account of (\ref{12})-(\ref{15}) and of the standard definition
\begin{equation}
\frac{d\sigma_{EW}(s,s')}{d\Omega}=
\frac{1}{16\pi^{2}}|M^{E}_{fi}+M^{W}_{fi}|^{2}
\label{16}
\end{equation}
can be presented as
$$d\sigma_{EW}^{A_{\nu_{l}}}(\theta_{EW},s,s')=
d\sigma_{E}^{A_{\nu_{l}}}(\theta_{E},s,s')+$$
\begin{equation}
+d\sigma_{I}^{A_{\nu_{l}}}(\theta_{I},s,s')+
d\sigma_{W}^{A_{\nu_{l}}}(\theta_{W},s,s'),
\label{17}
\end{equation}
in which $\theta_{EW}$ is the scattering angle of aa investigated neutrino
at the electroweakly united $(EW)$ emission of an axial-vector character.

To the purely Coulomb interaction answers the cross section
$$\frac{d\sigma_{E}^{A_{\nu_{l}}}(\theta_{E},s,s')}{d\Omega}=
\frac{1}{2}\sigma^{E}_{o}\{(1+ss')[g_{1\nu_{l}^{A}}^{E}-$$
$$-\frac{2}{3}<r^{2}_{\nu_{l}^{A}}>_{E}
(m_{\nu_{l}^{A}}^{E})^{2}\gamma_{E}^{-1}]^{2}+$$
\begin{equation}
+4(m_{\nu_{l}^{A}}^{E})^{2}\eta_{E}^{-2}(1-ss')g_{2\nu_{l}^{A}}^{2}
tg^{2}\frac{\theta_{E}}{2}\}F_{E}^{2}(q_{E}^{2}).
\label{18}
\end{equation}

The second cross section expresses an electroweak interference
$$\frac{d\sigma_{I}^{A_{\nu_{l}}}(\theta_{I},s,s')}{d\Omega}=
\frac{1}{2}\rho_{I}\sigma^{I}_{o}
g_{A_{\nu_{l}}}(1+ss')\{g_{1\nu_{l}^{A}}^{I}-$$
\begin{equation}
-\frac{2}{3}<r^{2}_{\nu_{l}^{A}}>_{I}(m_{\nu_{l}^{A}}^{I})^{2}
\gamma_{I}^{-1}\}F_{I}(q_{I}^{2}).
\label{19}
\end{equation}

The cross section of the weak interaction has the form
$$\frac{d\sigma_{W}^{A_{\nu_{l}}}(\theta_{W},s,s')}{d\Omega}=
\frac{G_{F}^{2}(m_{\nu_{l}^{A}}^{W})^{2}}{16\pi^{2}\sin^{2}\theta_{W}}
g_{A_{\nu_{l}}}^{2}\eta_{W}^{-2}(1+ss')[1-$$
\begin{equation}
-\eta_{W}^{2}]F_{W}^{2}(q_{W}^{2})
\cos^{2}\frac{\theta_{W}}{2}.
\label{20}
\end{equation}
Here we have also used the notations
$$\sigma_{o}^{E}=\frac{\alpha^{2}}{4(m_{\nu_{l}^{A}}^{E})^{2}}
\frac{\gamma_{E}^{2}}{\alpha_{E}}, \, \, \, \,
\rho_{I}=-\frac{2G_{F}(m_{\nu_{l}^{A}}^{I})^{2}}
{\pi\sqrt{2}\alpha\sin\theta_{W}}\gamma_{I}^{-1},$$
$$\sigma_{o}^{I}=\frac{\alpha^{2}}{4(m_{\nu_{l}^{A}}^{I})^{2}}
\frac{\gamma_{I}^{2}}{\alpha_{I}}, \, \, \, \,
\alpha_{K}=\frac{\eta^{2}_{K}}{(1-\eta^{2}_{K})
\cos^{2}(\theta_{K}/2)},$$
$$\gamma_{K}=\frac{\eta^{2}_{K}}{(1-\eta^{2}_{K})
\sin^{2}(\theta_{K}/2)}, \, \, \, \,
\eta_{K}=\frac{m_{\nu_{l}^{A}}^{K}}{E_{\nu_{l}^{A}}^{K}},$$
$$F_{E}(q_{E}^{2})=ZF_{c}(q_{E}^{2}), \, \, \, \,
F_{I}(q_{I}^{2})=ZZ_{W}F_{c}^{2}(q_{I}^{2}),$$
$$F_{W}(q_{W}^{2})=Z_{W}F_{c}(q_{W}^{2}), \, \, \, \,
q_{K}^{2}=-4(m_{\nu_{l}^{A}}^{K})^{2}\gamma_{K}^{-1},$$
$$Z_{W}=\frac{1}{2}\{\beta_{A}^{(0)}(Z+N)+\beta_{A}^{(1)}(Z-N)\},$$
$$A=Z+N, \, \, \, \, M_{T}=\frac{1}{2}(Z-N),$$
$$\beta_{A}^{(0)}=-2\sin^{2}\theta_{W},
\, \, \, \, \beta_{A}^{(1)}=\frac{1}{2}-2\sin^{2}\theta_{W},$$
$$g_{A_{\nu_{l}}}=-\frac{1}{2}, \, \, \, \, K=E,I,W.$$

Of them the Coulomb, weak and interference scattering angles of neutrinos
with the energies
$$E_{\nu_{l}^{A}}^{K}=\sqrt{p_{K}^{2}+(m_{\nu_{l}^{A}}^{K})^{2}}$$
equal to $\theta_{K},$ the functions $F_{c}(q_{K}^{2})$ characterize at
all three processes the charge $(F_{c}(0)=1)$ form factors of an axial-vector
nucleus having the isospin $T$ and its projection $M_{T},$ $\beta_{A}^{(0)}$
and $\beta_{A}^{(1)}$ correspond to the isoscalar and isovector part constants
of this nucleus C-antisymmetrical weak neutral current.

The self interference terms $(g_{i\nu_{l}^{A}}^{E})^{2}$ and
$<r_{\nu_{l}^{A}}^{4}>_{E}$ in (\ref{18}) describe the Coulomb scattering
of the left - or right-handed \cite{3} axial-vector dineutrinos of the Dirac
and Majorana nature
\begin{equation}
(\nu_{lL}^{A}, {\bar \nu_{lR}^{A}}), \, \, \, \,
(\nu_{lR}^{A}, {\bar \nu_{lL}^{A}}).
\label{21}
\end{equation}

As well as the contribution $g_{1\nu_{l}^{A}}^{E}<r_{\nu_{l}^{A}}^{2}>_{E}$ of
the mixed interference between the interactions with an axial-vector photon of
the anapole charge and its radius testifies in favor of such paraneutrinos.
Their scattering on nuclei can also be explained by the weak neutral currents.
In this situation, (\ref{20}) and its structural contributions
$g_{A_{\nu_{l}}}^{2}$ lead us to an implication that difermions
appear in the interaction type dependence. Therefore, the mixedly
interference terms $g_{A_{\nu_{l}}}g_{1\nu_{l}^{A}}^{I}$ and
$g_{A_{\nu_{l}}}<r_{\nu_{l}^{A}}^{2}>_{I}$ in (\ref{19}) may serve as
the confirmation of the existence of C-noninvariant paraneutrinos of
an electroweak nature.

Here it is essentially to note that (\ref{16}) redoubles the size of mixedly
interference contributions. But the number of paraneutrinos and of those
processes, in which they are present coincide. This allows to separate each
of the mixedly interference part of the scattering cross section into the two.

The formulas (\ref{18})-(\ref{20}) contain the terms $(1+ss')$ and $(1-ss')$
which there arise at the scattering with $(s'=s)$ or without $(s'=-s)$ change
of helicities of the left $(s=-1)$ - and right $(s=+1)$-handed neutrinos of
true neutrality. Therefore, their summation over $s',$ gives the possibility
to write (\ref{17}) in the form
$$d\sigma_{EW}^{A_{\nu_{l}}}(\theta_{EW},s)=
d\sigma_{E}^{A_{\nu_{l}}}(\theta_{E},s)+
\frac{1}{2}d\sigma_{I}^{A_{\nu_{l}}}(\theta_{I},s)+$$
\begin{equation}
+\frac{1}{2}d\sigma_{I}^{A_{\nu_{l}}}(\theta_{I},s)+
d\sigma_{W}^{A_{\nu_{l}}}(\theta_{W},s).
\label{22}
\end{equation}

The first cross section here behaves as
$$d\sigma_{E}^{A_{\nu_{l}}}(\theta_{E},s)=
d\sigma_{E}^{A_{\nu_{l}}}(\theta_{E},g_{1\nu_{l}^{A}}^{E},s)+
\frac{1}{2}d\sigma_{E}^{A_{\nu_{l}}}
(\theta_{E},g_{1\nu_{l}^{A}}^{E},<r^{2}_{\nu_{l}^{A}}>_{E},s)+$$
$$+\frac{1}{2}d\sigma_{E}^{A_{\nu_{l}}}
(\theta_{E},g_{1\nu_{l}^{A}}^{E},<r^{2}_{\nu_{l}^{A}}>_{E},s)+$$
\begin{equation}
+d\sigma_{E}^{A_{\nu_{l}}}(\theta_{E},<r^{2}_{\nu_{l}^{A}}>_{E},s)+
d\sigma_{E}^{A_{\nu_{l}}}(\theta_{E},g_{2\nu_{l}^{A}}^{E},s),
\label{23}
\end{equation}

\begin{equation}
\frac{d\sigma_{E}^{A_{\nu_{l}}}
(\theta_{E},g_{1\nu_{l}^{A}}^{E},s)}{d\Omega}=
\frac{d\sigma_{E}^{A_{\nu_{l}}}
(\theta_{E},g_{1\nu_{l}^{A}}^{E},s'=s)}{d\Omega}=
\sigma^{E}_{o}(g_{1\nu_{l}^{A}}^{E})^{2}F_{E}^{2}(q^{2}_{E}),
\label{24}
\end{equation}

$$\frac{d\sigma_{E}^{A_{\nu_{l}}}
(\theta_{E},g_{1\nu_{l}^{A}}^{E},<r^{2}_{\nu_{l}^{A}}>_{E},s)}
{d\Omega}=
\frac{d\sigma_{E}^{A_{\nu_{l}}}
(\theta_{E},g_{1\nu_{l}^{A}}^{E},<r^{2}_{\nu_{l}^{A}}>_{E},s'=s)}
{d\Omega}=$$
\begin{equation}
=-\frac{2}{3}(m_{\nu_{l}^{A}}^{E})^{2}\gamma_{E}^{-1}\sigma^{E}_{o}
g_{1\nu_{l}^{A}}^{E}<r^{2}_{\nu_{l}^{A}}>_{E}F_{E}^{2}(q^{2}_{E}),
\label{25}
\end{equation}

$$\frac{d\sigma_{E}^{A_{\nu_{l}}}
(\theta_{E},<r^{2}_{\nu_{l}^{A}}>_{E},s)}{d\Omega}=
\frac{d\sigma_{E}^{A_{\nu_{l}}}
(\theta_{E},<r^{2}_{\nu_{l}^{A}}>_{E},s'=s)}{d\Omega}=$$
\begin{equation}
=\frac{4}{9}(m_{\nu_{l}^{A}}^{E})^{4}\gamma_{E}^{-2}\sigma^{E}_{o}
<r^{4}_{\nu_{l}^{A}}>_{E}F_{E}^{2}(q^{2}_{E}),
\label{26}
\end{equation}

$$\frac{d\sigma_{E}^{A_{\nu_{l}}}
(\theta_{E},g_{2\nu_{l}^{A}}^{E},s)}{d\Omega}=
\frac{d\sigma_{E}^{A_{\nu_{l}}}
(\theta_{E},g_{2\nu_{l}^{A}}^{E},s'=-s)}{d\Omega}=$$
\begin{equation}
=4(m_{\nu_{l}^{A}}^{E})^{2}\eta^{-2}_{E}\sigma^{E}_{o}
(g_{2\nu_{l}^{A}}^{E})^{2}F_{E}^{2}(q^{2}_{E})tg^{2}\frac{\theta_{E}}{2}.
\label{27}
\end{equation}

An electroweak interference contribution is equal to
$$d\sigma_{I}^{A_{\nu_{l}}}(\theta_{I},s)=
d\sigma_{I}^{A_{\nu_{l}}}
(\theta_{I},g_{A_{\nu_{l}}},g_{1\nu_{l}^{A}}^{I},s)+$$
\begin{equation}
+d\sigma_{I}^{A_{\nu_{l}}}
(\theta_{I},g_{A_{\nu_{l}}},<r^{2}_{\nu_{l}^{A}}>_{I},s),
\label{28}
\end{equation}

$$\frac{d\sigma_{I}^{A_{\nu_{l}}}
(\theta_{I},g_{A_{\nu_{l}}},g_{1\nu_{l}^{A}}^{I},s)}{d\Omega}=
\frac{d\sigma_{I}^{A_{\nu_{l}}}
(\theta_{I},g_{A_{\nu_{l}}},g_{1\nu_{l}^{A}}^{I},s'=s)}
{d\Omega}=$$
\begin{equation}
=\rho_{I}\sigma^{I}_{o}g_{A_{\nu_{l}}}g_{1\nu_{l}^{A}}^{I}F_{I}(q_{I}^{2}),
\label{29}
\end{equation}

$$\frac{d\sigma_{I}^{A_{\nu_{l}}}
(\theta_{I},g_{A_{\nu_{l}}},<r^{2}_{\nu_{l}^{A}}>_{I},s)}
{d\Omega}=
\frac{d\sigma_{I}^{A_{\nu_{l}}}
(\theta_{I},g_{A_{\nu_{l}}},<r^{2}_{\nu_{l}^{A}}>_{I},s'=s)}
{d\Omega}+$$
\begin{equation}
=-\frac{2}{3}(m_{\nu_{l}^{A}}^{I})^{2}\gamma_{I}^{-1}
\rho_{I}\sigma^{I}_{o}
g_{A_{\nu_{l}}}<r^{2}_{\nu_{l}^{A}}>_{I}F_{I}(q_{I}^{2}).
\label{30}
\end{equation}

Third cross section has the following structure:
$$\frac{d\sigma_{W}^{A_{\nu_{l}}}
(\theta_{W},g_{A_{\nu_{l}}},s)}{d\Omega}=
\frac{d\sigma_{W}^{A_{\nu_{l}}}
(\theta_{W},g_{A_{\nu_{l}}},s'=s)}{d\Omega}=$$
\begin{equation}
=\frac{G_{F}^{2}(m_{\nu_{l}^{A}}^{W})^{2}}{8\pi^{2}\sin^{2}\theta_{W}}
g_{A_{\nu_{l}}}^{2}\eta_{W}^{-2}(1-\eta_{W}^{2})
F_{W}^{2}(q_{W}^{2})\cos^{2}\frac{\theta_{W}}{2}.
\label{31}
\end{equation}

Among (\ref{22})-(\ref{31}) the formulas (\ref{25}) and (\ref{30})
distinguish from others by the negative signs. In this appears a latent
connection between the anapole $g_{1\nu_{l}^{A}}(0)$ charge of the neutrino
and its charge $r_{\nu_{l}^{A}}$ radius. On the other hand, as was mentioned
earlier \cite{4,7}, the coexistence of $g_{1\nu_{l}^{A}}$ and
$g_{2\nu_{l}^{A}}$ is, by itself, not excluded. Therefore, in the presence of
an axial-vector mass, any truly neutral neutrino must possess simultaneously
each of all types of $A_{\nu_{l}}$ currents.

If now average the cross sections (\ref{18})-(\ref{20}) over $s$ and sum
over $s',$ one can reduce (\ref{17}) to the form
$$d\sigma_{EW}^{A_{\nu_{l}}}(\theta_{EW})=
d\sigma_{E}^{A_{\nu_{l}}}(\theta_{E})+
\frac{1}{2}d\sigma_{I}^{A_{\nu_{l}}}(\theta_{I})+$$
\begin{equation}
+\frac{1}{2}d\sigma_{I}^{A_{\nu_{l}}}(\theta_{I})+
d\sigma_{W}^{A_{\nu_{l}}}(\theta_{W}).
\label{32}
\end{equation}

Its structural parts may be written as
$$d\sigma_{E}^{A_{\nu_{l}}}(\theta_{E})=
d\sigma_{E}^{A_{\nu_{l}}}(\theta_{E},g_{1\nu_{l}^{A}}^{E})+
\frac{1}{2}
d\sigma_{E}^{A_{\nu_{l}}}
(\theta_{E},g_{1\nu_{l}^{A}}^{E},<r^{2}_{\nu_{l}^{A}}>_{E})+$$
$$+\frac{1}{2}
d\sigma_{E}^{A_{\nu_{l}}}
(\theta_{E},g_{1\nu_{l}^{A}}^{E},<r^{2}_{\nu_{l}^{A}}>_{E})+$$
\begin{equation}
+d\sigma_{E}^{A_{\nu_{l}}}(\theta_{E},<r^{2}_{\nu_{l}^{A}}>_{E})+
d\sigma_{E}^{A_{\nu_{l}}}(\theta_{E},g_{2\nu_{l}^{A}}^{E}),
\label{33}
\end{equation}
\begin{equation}
d\sigma_{I}^{A_{\nu_{l}}}(\theta_{I})=
d\sigma_{I}^{A_{\nu_{l}}}
(\theta_{I},g_{A_{\nu_{l}}},g_{1\nu_{l}^{A}}^{I})+
d\sigma_{I}^{A_{\nu_{l}}}
(\theta_{I},g_{A_{\nu_{l}}},<r^{2}_{\nu_{l}^{A}}>_{I}),
\label{34}
\end{equation}
\begin{equation}
d\sigma_{W}^{A_{\nu_{l}}}(\theta_{W})=
d\sigma_{W}^{A_{\nu_{l}}}(\theta_{W},g_{A_{\nu_{l}}}).
\label{35}
\end{equation}

Any component of each of (\ref{33})-(\ref{35}) has the same value as the
corresponding contribution from (\ref{23}), (\ref{28}), (\ref{31}) and that,
consequently, there exist the connections
\begin{equation}
d\sigma_{E}^{A_{\nu_{l}}}(\theta_{E},g_{1\nu_{l}^{A}}^{E})=
d\sigma_{E}^{A_{\nu_{l}}}(\theta_{E},g_{1\nu_{l}^{A}}^{E},s),
\label{36}
\end{equation}
\begin{equation}
d\sigma_{E}^{A_{\nu_{l}}}
(\theta_{E},g_{1\nu_{l}^{A}}^{E},<r^{2}_{\nu_{l}^{A}}>_{E})=
d\sigma_{E}^{A_{\nu_{l}}}
(\theta_{E},g_{1\nu_{l}^{A}}^{E},<r^{2}_{\nu_{l}^{A}}>_{E},s),
\label{37}
\end{equation}
\begin{equation}
d\sigma_{E}^{A_{\nu_{l}}}(\theta_{E},<r^{2}_{\nu_{l}^{A}}>_{E})=
d\sigma_{E}^{A_{\nu_{l}}}(\theta_{E},<r^{2}_{\nu_{l}^{A}}>_{E},s),
\label{38}
\end{equation}
\begin{equation}
d\sigma_{E}^{A_{\nu_{l}}}(\theta_{E},g_{2\nu_{l}^{A}}^{E})=
d\sigma_{E}^{A_{\nu_{l}}}(\theta_{E},g_{2\nu_{l}^{A}}^{E},s),
\label{39}
\end{equation}
\begin{equation}
d\sigma_{I}^{A_{\nu_{l}}}
(\theta_{I},g_{A_{\nu_{l}}},g_{1\nu_{l}^{A}}^{I})=
d\sigma_{I}^{A_{\nu_{l}}}
(\theta_{I},g_{A_{\nu_{l}}},g_{1\nu_{l}^{A}}^{I},s),
\label{40}
\end{equation}
\begin{equation}
d\sigma_{I}^{A_{\nu_{l}}}
(\theta_{I},g_{A_{\nu_{l}}},<r^{2}_{\nu_{l}^{A}}>_{I})=
d\sigma_{I}^{A_{\nu_{l}}}
(\theta_{I},g_{A_{\nu_{l}}},<r^{2}_{\nu_{l}^{A}}>_{I},s),
\label{41}
\end{equation}
\begin{equation}
d\sigma_{W}^{A_{\nu_{l}}}(\theta_{W},g_{A_{\nu_{l}}})=
d\sigma_{W}^{A_{\nu_{l}}}(\theta_{W},g_{A_{\nu_{l}}},s).
\label{42}
\end{equation}

It is clear from the above expansions (\ref{22}) and (\ref{32}) that
a flux of the scattered neutrinos becomes a partially ordered class of
outgoing axial-vector particles. They can therefore be described by the
partially ordered set of the corresponding cross sections:
\begin{equation}
d\sigma_{EW}^{A_{\nu_{l}}}=
\{d\sigma_{EW}^{A_{\nu_{l}}}(\theta_{EW},s), \, \, \, \,
d\sigma_{EW}^{A_{\nu_{l}}}(\theta_{EW})\}.
\label{43}
\end{equation}

We have already mentioned that any of (\ref{22}) and (\ref{32}) constitutes
a kind of the structural subclass:
$$d\sigma_{EW}^{A_{\nu_{l}}}(\theta_{EW},s)=
\{d\sigma_{E}^{A_{\nu_{l}}}(\theta_{E},g_{1\nu_{l}^{A}}^{E},s),  \, \, \, \,
\frac{1}{2}
d\sigma_{E}^{A_{\nu_{l}}}
(\theta_{E},g_{1\nu_{l}^{A}}^{E},<r^{2}_{\nu_{l}^{A}}>_{E},s),$$
$$\frac{1}{2}
d\sigma_{E}^{A_{\nu_{l}}}
(\theta_{E},g_{1\nu_{l}^{A}}^{E},<r^{2}_{\nu_{l}^{A}}>_{E},s),
\, \, \, \,
d\sigma_{E}^{A_{\nu_{l}}}(\theta_{E},<r^{2}_{\nu_{l}^{A}}>_{E},s),$$
$$d\sigma_{E}^{A_{\nu_{l}}}(\theta_{E},g_{2\nu_{l}^{A}}^{E},s),
\, \, \, \,
\frac{1}{2}
d\sigma_{I}^{A_{\nu_{l}}}
(\theta_{I},g_{A_{\nu_{l}}},g_{1\nu_{l}^{A}}^{I},s),$$
$$\frac{1}{2}
d\sigma_{I}^{A_{\nu_{l}}}
(\theta_{I},g_{A_{\nu_{l}}},g_{1\nu_{l}^{A}}^{I},s),
\, \, \, \,
\frac{1}{2}
d\sigma_{I}^{A_{\nu_{l}}}
(\theta_{I},g_{A_{\nu_{l}}},<r^{2}_{\nu_{l}^{A}}>_{I},s),$$
\begin{equation}
\frac{1}{2}
d\sigma_{I}^{A_{\nu_{l}}}
(\theta_{I},g_{A_{\nu_{l}}},<r^{2}_{\nu_{l}^{A}}>_{I},s),
\, \, \, \,
d\sigma_{W}^{A_{\nu_{l}}}(\theta_{W},g_{A_{\nu_{l}}},s)\},
\label{44}
\end{equation}
$$d\sigma_{EW}^{A_{\nu_{l}}}(\theta_{EW})=
\{d\sigma_{E}^{A_{\nu_{l}}}(\theta_{E},g_{1\nu_{l}^{A}}^{E}),
\, \, \, \,
\frac{1}{2}
d\sigma_{E}^{A_{\nu_{l}}}
(\theta_{E},g_{1\nu_{l}^{A}}^{E},<r^{2}_{\nu_{l}^{A}}>_{E}),$$
$$\frac{1}{2}
d\sigma_{E}^{A_{\nu_{l}}}
(\theta_{E},g_{1\nu_{l}^{A}}^{E},<r^{2}_{\nu_{l}^{A}}>_{E}),
\, \, \, \,
d\sigma_{E}^{A_{\nu_{l}}}(\theta_{E},<r^{2}_{\nu_{l}^{A}}>_{E}),$$
$$d\sigma_{E}^{A_{\nu_{l}}}(\theta_{E},g_{2\nu_{l}^{A}}^{E}), \, \, \, \,
\frac{1}{2}
d\sigma_{I}^{A_{\nu_{l}}}
(\theta_{I},g_{A_{\nu_{l}}},g_{1\nu_{l}^{A}}^{I}),$$
$$\frac{1}{2}
d\sigma_{I}^{A_{\nu_{l}}}
(\theta_{I},g_{A_{\nu_{l}}},g_{1\nu_{l}^{A}}^{I}),
\, \, \, \,
\frac{1}{2}
d\sigma_{I}^{A_{\nu_{l}}}
(\theta_{I},g_{A_{\nu_{l}}},<r^{2}_{\nu_{l}^{A}}>_{I}),$$
\begin{equation}
\frac{1}{2}
d\sigma_{I}^{A_{\nu_{l}}}
(\theta_{I},g_{A_{\nu_{l}}},<r^{2}_{\nu_{l}^{A}}>_{I}),
\, \, \, \,
d\sigma_{W}^{A_{\nu_{l}}}(\theta_{W},g_{A_{\nu_{l}}})\}.
\label{45}
\end{equation}

From the point of view of (\ref{36})-(\ref{42}), these subsets coincide.
However, an equality of cross sections (\ref{22}) and (\ref{32}) can exist
only in the case when all elements correspond in sets (\ref{44}) and (\ref{45})
to the same spin state of the above-mentioned paraneutrinos. Such a principle
can explain the identicality of the cross section structural components of
the discussed type of interaction.

It is also relevant to observe that between the two neutrinos of
each of parafermions (\ref{21}) there exists a flavor symmetrical
connection \cite{7}. This gives the possibility to understand the flavor
symmetry as a theorem \cite{3} about the equality of the structural parts
of the neutrino interaction cross sections with an electroweak field of
an axial-vector nature.

\begin{center}
{\bf 3. Mass structure of neutrinos of true neutrality}
\end{center}

According to the above reasoning, the ratio of any pair of elements in
(\ref{43}) is equal to unity. Similar connections establish a set of the
forty two equalities. It together with the values of cross sections
(\ref{36})-(\ref{42}) constitutes the system of the twenty one most
diverse equations.

To show their features, it is desirable to investigate the five relations
from the first-initial system:
\begin{equation}
\frac{d\sigma_{W}^{A_{\nu_{l}}}(\theta_{W},g_{A_{\nu_{l}}})}
{d\sigma_{E}^{A_{\nu_{l}}}(\theta_{E},g_{i\nu_{l}^{A}}^{E})}=1,
\, \, \, \,
\frac{2d\sigma_{W}^{A_{\nu_{l}}}(\theta_{W},g_{A_{\nu_{l}}})}
{d\sigma_{I}^{A_{\nu_{l}}}
(\theta_{I},g_{A_{\nu_{l}}},g_{1\nu_{l}^{A}}^{I})}=1,
\label{46}
\end{equation}
\begin{equation}
\frac{d\sigma_{W}^{A_{\nu_{l}}}(\theta_{W},g_{A_{\nu_{l}}})}
{d\sigma_{E}^{A_{\nu_{l}}}(\theta_{E},<r^{2}_{\nu_{l}^{A}}>_{E})}=1,
\, \, \, \,
\frac{2d\sigma_{W}^{A_{\nu_{l}}}(\theta_{W},g_{A_{\nu_{l}}})}
{d\sigma_{I}^{A_{\nu_{l}}}
(\theta_{I},g_{A_{\nu_{l}}},<r^{2}_{\nu_{l}^{A}}>_{I})}=1.
\label{47}
\end{equation}

It is also convenient to use the nucleus with zero spin and isospin only,
so as the target nucleus isotopic structure can essentially change the innate
properties of a neutrino \cite{18}. Therefore, if we suppose that $N=Z,$ a
unification of (\ref{46}) and (\ref{47}) with sizes of cross sections from
(\ref{36})-(\ref{42}) leads us to those equations which at the choice of a
particle energy $E_{\nu_{l}^{A}}^{K}\gg m_{\nu_{l}^{A}}^{K}$ when
$$lim_{\eta_{E}\rightarrow 0,\theta_{E}\rightarrow 0}
\frac{tg^{2}(\theta_{E}/2)}{\eta_{E}^{2}}=\frac{1}{4},$$
$$lim_{\eta_{K}\rightarrow 0,\theta_{K}\rightarrow 0}
\frac{\eta^{2}_{K}}{(1-\eta^{2}_{K})
\sin^{2}(\theta_{K}/2)}=-2$$
allow to establish here an explicit mass structure dependence of neutrino
currents of true neutrality
\begin{equation}
g_{1\nu_{l}^{A}}^{E}(0)=-g_{A_{\nu_{l}}}
\frac{G_{F}m_{\nu_{l}^{A}}^{E}m_{\nu_{l}^{A}}^{W}}
{\pi\sqrt{2}\alpha}\sin\theta_{W},
\label{48}
\end{equation}
\begin{equation}
g_{2\nu_{l}^{A}}^{E}(0)=-g_{A_{\nu_{l}}}
\frac{G_{F}m_{\nu_{l}^{A}}^{W}}{\pi\sqrt{2}\alpha}\sin\theta_{W},
\label{49}
\end{equation}
\begin{equation}
<r^{2}_{\nu_{l}^{A}}>_{E}=-g_{A_{\nu_{l}}}
\frac{3G_{F}}{\pi\sqrt{2}\alpha}
\frac{m_{\nu_{l}^{A}}^{W}}{m_{\nu_{l}^{A}}^{E}}\sin\theta_{W},
\label{50}
\end{equation}
\begin{equation}
g_{1\nu_{l}^{A}}^{I}(0)=-g_{A_{\nu_{l}}}
\frac{G_{F}(m_{\nu_{l}^{A}}^{W})^{2}}{\pi\sqrt{2}\alpha}\sin\theta_{W},
\label{51}
\end{equation}
\begin{equation}
<r^{2}_{\nu_{l}^{A}}>_{I}=-g_{A_{\nu_{l}}}
\frac{3G_{F}}{\pi\sqrt{2}\alpha}
\left(\frac{m_{\nu_{l}^{A}}^{W}}{m_{\nu_{l}^{A}}^{I}}\right)^{2}\sin\theta_{W}.
\label{52}
\end{equation}

Thus, on the basis of
\begin{equation}
e_{\nu_{l}^{A}}^{E}=-g_{A_{\nu_{l}}}
\frac{G_{F}m_{\nu_{l}^{A}}^{E}m_{\nu_{l}^{A}}^{W}}
{\pi\sqrt{2}\alpha}\sin\theta_{W},
\label{53}
\end{equation}
we can relate with confidence each of (\ref{48})-(\ref{52}) to a renormalized
value of a neutrino C-noninvariant electric charge. In this appears the chance
for their presentation in a latent united form
\begin{equation}
g_{1\nu_{l}^{A}}^{E}(0)=e_{\nu_{l}^{A}}^{E},
\label{54}
\end{equation}
\begin{equation}
g_{2\nu_{l}^{A}}^{E}(0)=\frac{e_{\nu_{l}^{A}}^{E}}{m_{\nu_{l}^{A}}^{E}},
\label{55}
\end{equation}
\begin{equation}
<r^{2}_{\nu_{l}^{A}}>_{E}=
\frac{3e_{\nu_{l}^{A}}^{E}}{(m_{\nu_{l}^{A}}^{E})^{2}},
\label{56}
\end{equation}
\begin{equation}
g_{1\nu_{l}^{A}}^{I}(0)=
\frac{m_{\nu_{l}^{A}}^{W}}{m_{\nu_{l}^{A}}^{E}}e_{\nu_{l}^{A}}^{E},
\label{57}
\end{equation}
\begin{equation}
<r^{2}_{\nu_{l}^{A}}>_{I}=
\frac{m_{\nu_{l}^{A}}^{E}m_{\nu_{l}^{A}}^{W}}
{(m_{\nu_{l}^{A}}^{I})^{2}}<r^{2}_{\nu_{l}^{A}}>_{E}.
\label{58}
\end{equation}

It is seen from these currents that $g_{2\nu_{l}^{A}}^{E}$ gives the normal
Dirac value of an electric dipole moment of neutrino. Its anomalous part
can therefore be differ from zero only if an interaction there arises at
the expense of exchange by the two bosons.

\begin{center}
{\bf 4. Anomalous behavior of neutrinos of axial-vector currents}
\end{center}

It is not excluded, however, that the dimensionality of the normal
values of currents $e_{\nu_{l}^{A}}^{norm},$ $d_{\nu_{l}^{A}}^{norm}$ and
$<r^{2}_{\nu_{l}^{A}}>_{norm}$ can be established by the following manner:
$e_{\nu_{l}^{A}}^{norm}=eg_{1\nu_{l}^{A}}^{E}(0),$
$d_{\nu_{l}^{A}}^{norm}=eg_{2\nu_{l}^{A}}^{E}(0),$
$<r^{2}_{\nu_{l}^{A}}>_{norm}=<r^{2}_{\nu_{l}^{A}}>_{E}.$

Therefore, from the point of view of the Schwinger size of the anomalous
magnetic moment \cite{19}, it should be expected that $d_{\nu_{l}^{A}}^{anom}=
(\alpha/2\pi)d_{\nu_{l}^{A}}^{norm}.$ Then it is possible, for example, to
present the functions (\ref{48})-(\ref{50}) in the form
\begin{equation}
e_{\nu_{l}^{A}}^{anom}=-g_{A_{\nu_{l}}}
\frac{eG_{F}m_{\nu_{l}^{A}}^{E}m_{\nu_{l}^{A}}^{W}}
{2\pi^{2}\sqrt{2}}\sin\theta_{W},
\label{59}
\end{equation}
\begin{equation}
d_{\nu_{l}^{A}}^{anom}=-g_{A_{\nu_{l}}}
\frac{eG_{F}m_{\nu_{l}^{A}}^{W}}{2\pi^{2}\sqrt{2}}\sin\theta_{W},
\label{60}
\end{equation}
\begin{equation}
<r^{2}_{\nu_{l}^{A}}>_{anom}=-g_{A_{\nu_{l}}}
\frac{3G_{F}}{2\pi^{2}\sqrt{2}}
\frac{m_{\nu_{l}^{A}}^{W}}{m_{\nu_{l}^{A}}^{E}}\sin\theta_{W},
\label{61}
\end{equation}

They together with $e_{\nu_{l}^{A}}^{norm},$ $d_{\nu_{l}^{A}}^{norm}$ and
$<r^{2}_{\nu_{l}^{A}}>_{norm}$ define the full anapole and its radius as
well as an electric dipole: $e_{\nu_{l}^{A}}^{full}=
(1+\alpha/2\pi)e_{\nu_{l}^{A}}^{norm},$ $d_{\nu_{l}^{A}}^{full}=
(1+\alpha/2\pi)d_{\nu_{l}^{A}}^{norm},$ $<r^{2}_{\nu_{l}^{A}}>_{full}=
(1+\alpha/2\pi)<r^{2}_{\nu_{l}^{A}}>_{norm}.$

Such an approach is based actually on the appearance of
$d_{\nu_{l}^{A}}^{anom}$ in the anomalous C-odd electric charge
\cite{7,11} dependence.

It is also relevant to include in the discussion the earlier known case when
$m_{\nu_{l}^{A}}^{E}=m_{\nu_{l}^{A}}^{W}=m_{\nu_{l}^{A}}.$ This circumstance
replaces (\ref{59})-(\ref{61}) for
\begin{equation}
e_{\nu_{l}^{A}}^{anom}=-g_{A_{\nu_{l}}}
\frac{eG_{F}m_{\nu_{l}^{A}}^{2}}{2\pi^{2}\sqrt{2}}\sin\theta_{W},
\label{62}
\end{equation}
\begin{equation}
d_{\nu_{l}^{A}}^{anom}=-g_{A_{\nu_{l}}}
\frac{eG_{F}m_{\nu_{l}^{A}}}{2\pi^{2}\sqrt{2}}\sin\theta_{W},
\label{63}
\end{equation}
\begin{equation}
<r^{2}_{\nu_{l}^{A}}>_{anom}=-g_{A_{\nu_{l}}}
\frac{3G_{F}}{2\pi^{2}\sqrt{2}}\sin\theta_{W},
\label{64}
\end{equation}

Furthermore, if we accept that $g_{A_{\nu_{l}}}=1,$ $\beta_{A}^{(0)}=1$ then
(\ref{46}) and (\ref{47}) transform (\ref{59})-(\ref{61}) into the following:
\begin{equation}
e_{\nu_{l}^{A}}^{anom}=
\frac{eG_{F}m_{\nu_{l}^{A}}^{2}}{4\pi^{2}\sqrt{2}},
\label{65}
\end{equation}
\begin{equation}
d_{\nu_{l}^{A}}^{anom}=
\frac{eG_{F}m_{\nu_{l}^{A}}}{4\pi^{2}\sqrt{2}},
\label{66}
\end{equation}
\begin{equation}
<r^{2}_{\nu_{l}^{A}}>_{anom}=
\frac{3G_{F}}{4\pi^{2}\sqrt{2}}
\label{67}
\end{equation}
which are valid at the level of a universal $(V-A)$ theory \cite{20,21}.

\begin{center}
{\bf 5. An axial-vector neutrino universality}
\end{center}

Our equations (\ref{48})-(\ref{58}) have a generality for all types of
leptons of an axial-vector nature. We can, therefore, relate, for example,
(\ref{48})-(\ref{50}) to a renormalized size of an electron C-noninvariant
electric charge \cite{22}
\begin{equation}
e_{e^{A}}^{E}=-g_{A_{e}}
\frac{G_{F}m_{e^{A}}^{E}m_{e^{A}}^{W}}{\pi\sqrt{2}\alpha}\sin\theta_{W}.
\label{68}
\end{equation}

This indicates to that at $g_{A_{e}}=g_{A_{\nu_{l}}}$ we find
\begin{equation}
e_{\nu_{l}^{A}}^{E}=g_{1\nu_{l}^{A}}^{E}(0)=
\frac{m_{\nu_{l}^{A}}^{E}}{m_{e^{A}}^{E}}
\frac{m_{\nu_{l}^{A}}^{W} }{m_{e^{A}}^{W}}e_{e^{A}}^{E},
\label{69}
\end{equation}
\begin{equation}
d_{\nu_{l}^{A}}^{E}=
g_{2\nu_{l}^{A}}^{E}(0)=\frac{m_{\nu_{l}^{A}}^{W}}{m_{e^{A}}^{W}}
\frac{e_{e^{A}}^{E}}{m_{e^{A}}^{E}},
\label{70}
\end{equation}
\begin{equation}
<r^{2}_{\nu_{l}^{A}}>_{E}=
\frac{m_{\nu_{l}^{A}}^{W}}{m_{e^{A}}^{W}}
\frac{3e_{e^{A}}^{E}}{m_{\nu_{l}^{A}}^{E}m_{e^{A}}^{E}}.
\label{71}
\end{equation}

Exactly the same one can apply once again to (\ref{48}) and (\ref{49}),
because they predict a connection between the masses of each lepton
and its neutrino:
\begin{equation}
\frac{m_{\nu_{l}^{A}}^{E}}{m_{l^{A}}^{E}}=
\frac{g_{1\nu_{l}^{A}}^{E}(0)}{g_{1l^{A}}^{E}(0)}
\frac{g_{2l^{A}}^{E}(0)}{g_{2\nu_{l}^{A}}^{E}(0)}.
\label{72}
\end{equation}

The latter implies that a full true number conservation (\ref{7}) suggests
the following value of an axial-vector neutrino electric mass
\begin{equation}
m_{\nu_{e}^{A}}^{E}:m_{\nu_{\mu}^{A}}^{E}:m_{\nu_{\tau}^{A}}^{E}=
m_{e^{A}}^{E}:m_{\mu^{A}}^{E}:m_{\tau^{A}}^{E}.
\label{73}
\end{equation}

In a similar way one can get from (\ref{49}) and (\ref{51}) the size
of the weak masses of truly neutral types of neutrinos
\begin{equation}
m_{\nu_{e}^{A}}^{W}:m_{\nu_{\mu}^{A}}^{W}:m_{\nu_{\tau}^{A}}^{W}=
m_{e^{A}}^{W}:m_{\mu^{A}}^{W}:m_{\tau^{A}}^{W}.
\label{74}
\end{equation}

These values together with (\ref{73}) require the existence of a connection
\begin{equation}
m_{\nu_{e}^{A}}^{E}m_{\nu_{e}^{A}}^{W}:
m_{\nu_{\mu}^{A}}^{E}m_{\nu_{\mu}^{A}}^{W}:
m_{\nu_{\tau}^{A}}^{E}m_{\nu_{\tau}^{A}}^{W}=
m_{e^{A}}^{E}m_{e^{A}}^{W}:
m_{\mu^{A}}^{E}m_{\mu^{A}}^{W}:
m_{\tau^{A}}^{E}m_{\tau^{A}}^{W}.
\label{75}
\end{equation}

Therefore, without contradicting ideas of lepton universality \cite{23}
\begin{equation}
m_{l^{A}}^{E}m_{l^{A}}^{W}=const,
\label{76}
\end{equation}
we not only recognize that
\begin{equation}
m_{\nu_{l}^{A}}^{E}m_{\nu_{l}^{A}}^{W}=const
\label{77}
\end{equation}
but also need to elucidate the nature of axial-vector types of neutrinos
from the point of view of a neutrino universality principle.

\begin{center}
{\bf 6. Conclusion}
\end{center}

Analysis of all known experiments about a neutrino axial-vector
radius is, at present, based simply on the assumption \cite{24} that
$<r^{2}_{\nu_{l}^{A}}>_{E}=-6G_{1\nu_{l}^{A}}(0).$ This implication, as we
have seen above, does not correspond to the reality. Thereby, the possibility
to compare $<r^{2}_{\nu_{l}^{A}}>_{E}$ with an experiment is absent.

There exist, however, earlier laboratory and cosmological limits for
the neutrino electric dipole moment. The first of them \cite{15} corresponds
to the light neutrino of C-odd character and is equal \cite{25} to
\begin{equation}
d_{\nu_{e}^{A}}^{E}< 0,44\cdot10^{-20}\ {\rm e\cdot cm}.
\label{78}
\end{equation}

The second refers to any type of the C-noninvariant neutrino \cite{16}
of a Dirac nature:
\begin{equation}
d_{\nu_{l}^{A}}^{E}< 2,5\cdot 10^{-22}\ {\rm e\cdot cm}.
\label{79}
\end{equation}

In both restrictions, $e$ coincides with an electric $e_{l^{V}}^{E}$ charge
of lepton of the vector type. A unification of (\ref{55}), (\ref{78}) and
(\ref{79}) at such a situation establishes those values of the neutrino
electric masses which distinguish from the available data \cite{26} in
the literature. This becomes possible owing to the inequalities
\begin{equation}
e_{\nu_{l}^{A}}^{K}\neq e_{\nu_{l}^{V}}^{K}, \, \, \, \,
m_{\nu_{l}^{A}}^{K}\neq m_{\nu_{l}^{V}}^{K}.
\label{80}
\end{equation}

Their existence, however, does not contradict our observation that truly
neutral neutrinos possessing the mass of an axial-vector character \cite{7}
have no vector interactions \cite{3}.

Returning to (\ref{54}) and (\ref{55}), we remark that they give the size
of an electric mass of the C-odd neutrino
\begin{equation}
\frac{g_{1\nu_{l}^{A}}^{E}(0)}{g_{2\nu_{l}^{A}}^{E}(0)}=m_{\nu_{l}^{A}}^{E}.
\label{81}
\end{equation}

Its comparison with the ratio of currents (\ref{55}) and (\ref{57})
establishes (\ref{77}) and thereby confirms the fact that all neutrinos
of true neutrality regardless of the difference in masses of an axial-vector
nature, possess the same anapole charge with his radius as well as an equal
electric dipole moment.

On the other hand, as was noted in the work \cite{22}, the standard
formula \cite{23} for $e_{l^{V}}^{E}$ at the availability of a connection
\begin{equation}
\frac{G_{F}}{\pi\sqrt{2}\alpha}=
\frac{1}{2m_{W}^{2}}\frac{1}{\sin^{2}\theta_{W}}
\label{82}
\end{equation}
has the following form
\begin{equation}
e_{l^{V}}^{E}=-g_{V_{l}}\frac{m_{l^{V}}^{E}m_{l^{V}}^{W}}
{2m_{W^{\pm}}^{2}}\frac{1}{\sin\theta_{W}}.
\label{83}
\end{equation}

This would seem to say about that (\ref{82}) allows to express the function
(\ref{48}) through the C-invariant charge of lepton of a vector nature. It
is easy to see, however, that a unification of (\ref{48}), (\ref{49}) and
(\ref{83}) jointly with (\ref{78}) and (\ref{79}) gives the possibility to
define the other values of the weak axial-vector masses instead of their
innate sizes. They state that the C-odd neutrinos have no interactions
with $W^{\pm}$-bosons but possess with $W^{0}$-bosons the same interaction
as the nucleons of true neutrality.

Therefore, we present (\ref{53}) in the form
\begin{equation}
e_{\nu_{l}^{A}}^{E}=-g_{A_{\nu_{l}}}
\frac{m_{\nu_{l}^{A}}^{E}m_{\nu_{l}^{A}}^{W}}
{2m_{W^{0}}^{2}}\frac{1}{\sin\theta_{W}}.
\label{84}
\end{equation}

According to these results, each C-even or C-odd neutrino may serve as the
source, respectively, of a vector or an axial-vector intermediate boson.

So, it is seen that a neutrino itself testifies in favor of the
existence of massive $Z^{\pm} (W^{0})$-bosons. Their nature has been created
so that any type of the neutrino C-symmetrical (C-antisymmetrical) interaction
with field of emission originates at the expense of a kind of charged
(truly neutral) gauge boson. Among such particles one can find both vector
(axial-vector) photons and $Z^{\pm} (Z^{0})$-bosons. Under these circumstances,
the masses of bosons $Z^{0}(W^{0})$ and $Z^{\pm}(W^{\pm})$ are strictly
different
\begin{equation}
m_{Z^{0}}\neq m_{Z^{\pm}},
\label{85}
\end{equation}
\begin{equation}
m_{W^{0}}\neq m_{W^{\pm}}.
\label{86}
\end{equation}

In conformity with such implications of the standard electroweak
theory, we conclude \cite{27} that
\begin{equation}
m_{W^{0}}=m_{Z^{0}}\cos\theta_{W}.
\label{87}
\end{equation}
\begin{equation}
m_{W^{\pm}}=m_{Z^{\pm}}\cos\theta_{W}.
\label{88}
\end{equation}

Thus, all earlier establish properties of bosons $Z (W)$ can be
accepted as those characteristic features of fields which refer undoubtedly
only to vector particles $Z^{\pm} (W^{\pm})$ corresponding in nature to
axial-vector $Z^{0} (W^{0})$-bosons. They are of course united in triplets
$(Z^{-},$ $Z^{0},$ $Z^{+})$ and $(W^{-},$ $W^{0},$ $W^{+})$ because of
a hard connection of their masses.

This convinces us here that although suggested triboson unification requires
special verification, the above-mentioned inequalities (\ref{80}), (\ref{86})
and (\ref{87}) give the possibility to interpret not only the existence
of truly neutral types of Dirac neutrinos and gauge fields but also the
availability of mass structure in them as the one of known laboratory facts.

\end{document}